%% file: tardis-short.tex
\documentclass[usenatbib]{mnras}
\usepackage[utf8]{inputenc}
\usepackage{hyperref}
\usepackage{physics}
\newenvironment{comment}[1]{}{}

\usepackage{xspace}
\usepackage{tcolorbox}

\usepackage{tikz}
\let\plate\relax 
\usetikzlibrary{bayesnetWLee}

\usepackage[colorinlistoftodos]{todonotes} 

\usepackage{bm,amssymb,amsfonts,amsmath}  
\usepackage{graphicx}    
\graphicspath{
  {./figures/}
}
\newcommand{\lleq}[1]{\label{#1} }

\input{symbols}

\newcommand{\cond}{\,|\,}
\newcommand{\refeq}[1]{Equation~\ref{#1}}
\newcommand{\reffig}[1]{Figure~\ref{fig:#1}}
\newcommand{\refsec}[1]{Section~\ref{sec:#1}}
\newcommand{\reftab}[1]{\autoref{tab:#1}} 

\newcommand{\doroverline}[2]{\overline{#1#2}}
\newcommand{\roverline}[1]{\mathpalette\doroverline{#1}}

\newcommand{\CPD}{compound Poisson distribution\xspace}
\renewcommand{\CPD}{CPD\xspace}
\newcommand{\data}{\mathcal{D}}

\DeclareMathOperator{\Expect}{\mathbb{E}}
\newcommand{\expect}[1]{\Expect\left[#1\right]}

\DeclareMathOperator{\GammaDist}{Gamma}
\DeclareMathOperator{\GaussianDist}{\mathcal{N}}
\newcommand{\half}{\tfrac{1}{2}}
\DeclareMathOperator{\InvGammaDist}{InvGamma}

\newcommand{\npack}{{N_p}}
\newcommand{\Lumtot}{L}
\newcommand{\Lumuniq}{\Lumtot_{\infty}}

\newcommand{\Lumobs}{\Lumtot_{\rm obs}}
\newcommand{\lum}{\ell}
\newcommand{\Lum}{\lum}

\newcommand{\physpar}{\bmtheta}

\newcommand{\rmdx}[1]{\dd{#1}} 
\newcommand{\samplel}{\langle\lum\rangle}
\renewcommand{\samplel}{\roverline{\lum}}
\newcommand{\sampleltwo}{\ensuremath{\roverline{\lum^2}}}
\newcommand{\samplelvar}{\roverline{\sigma^2}}
\newcommand{\sigobs}{\sigma_{\rm obs}}
\newcommand{\varobs}{\sigobs^2}
\newcommand{\tardis}{\textsc{tardis}\xspace}
\DeclareMathOperator{\Variance}{\mathbb{V}}
\newcommand{\variance}[1]{\Variance\left[#1\right]}

\date{Accepted 2018 March 26. Received 2018 March 21; in original form 2017 May 30}

\title[Radiative-transfer statistical uncertainties]{Bayesian modelling of uncertainties of Monte Carlo radiative-transfer simulations } %
\author[F.~Beaujean et al.]{Frederik~Beaujean,$^{1,2}$
  Hans~C.~Eggers,$^{3,4}$ and
  Wolfgang~E.~Kerzendorf$^{1,5}$\\
  $^{1}$Excellence Cluster Universe, Boltzmannstraße 2, 85748 Garching bei München, Germany\\
  $^{2}$Fakultät für Physik, Ludwig-Maximilians-Universität München,
  Schellingstraße 4, 80799 München, Germany\\
  $^{3}$Institute of Theoretical Physics, Department of Physics, Stellenbosch
  University, P/Bag X1, 7602 Matieland, South Africa\\
  $^{4}$National Institute of Theoretical Physics, Stellenbosch, P/Bag X1, 7602 Matieland, South Africa \\
  $^{5}$European Southern Observatory, Karl-Schwarzschild-Straße 2, 85748
  Garching bei München, Germany\\
}

\begin{document}

\maketitle

\begin{abstract}
  One of the big challenges in astrophysics is the comparison of complex
  simulations to observations. As many codes do not directly generate
  observables (e.g. hydrodynamic simulations), the last step in the modelling
  process is often a radiative-transfer treatment. For this step, the community
  relies increasingly on Monte Carlo radiative transfer due to the ease of
  implementation and scalability with computing power.
  We consider simulations in which the number of photon packets is Poisson
  distributed, while the weight assigned to a single photon packet follows any
  distribution of choice. We show how to estimate the statistical uncertainty of
  the sum of weights in each bin from the output of a single radiative-transfer
  simulation.
  Our Bayesian approach produces a posterior distribution that is valid for any
  number of packets in a bin, even zero packets, and is easy to implement in
  practice. Our analytic results for large number of packets show that we
  generalise existing methods that are valid only in limiting cases. The
  statistical problem considered here appears in identical form in a wide range
  of Monte Carlo simulations including particle physics and importance sampling.
  It is particularly powerful in extracting information when the available data
  are sparse or quantities are small.
\end{abstract}


\section{Introduction} \label{sec:intro}

The role of astrophysics is to understand the existence and evolution of
physical objects in the framework of the fundamental physical laws that govern
our reality. The vast majority of astrophysical data comes in the form of
electromagnetic radiation constituting the final observable of highly complex
physical processes. If no analytic estimates are available, a common approach is
to construct a computer simulation to infer physics parameters by comparing the
simulation's synthetic observables to physical observables by some criterion or
metric.

Defining a metric that reliably indicates whether any given synthetic observable
is consistent with the measured data is a nontrivial task because both
measurement as well as simulation have associated uncertainties that need to be
taken into account. Uncertainty in the measurement itself results from
systematic effects, photon, and detector noise. In simulations, the dominant
uncertainty is often the choice of model or approximations that are needed to
get an answer in reasonable time.
Specifically for Monte Carlo simulations there is additional
uncertainty due to Monte Carlo noise.  Capturing this statistical
uncertainty is an important step towards generating a useful
comparison metric between experiment and simulation.

One major application for Monte Carlo simulation is radiative transfer where the
Monte Carlo approach
%
%
easily allows the implementation of complex microphysics in an era of vast
computational resources. This, however, results in synthetic observables that
exhibit Poisson noise coupled with other sources of noise. While the noise
distribution from actual observations is well understood, the noise from Monte
Carlo simulations is --- if at all --- only crudely estimated.

In this work, we focus on Monte Carlo radiative-transfer simulations in which
the fundamental unit is a photon packet with a frequency and a weight (energy,
luminosity \dots). As a concrete example, we use
\tardis\citep{2014MNRAS.440..387K}. \tardis\ approximates the propagation of
photons through supernova ejecta using a total of $\npack$ Monte Carlo photon
packets drawn from a black-body spectrum. As packets propagate through the
supernova envelope, they may change both frequency and luminosity and thus
approximate the effect of various scattering processes, local Doppler shifts,
and other radiative processes. The \tardis\ output then consists of $\npack$
packets with a frequency and a luminosity.

In the usual approach, the ensemble of packets is used to generate a synthetic
spectrum by adding the observed luminosities of all packets in a given frequency
bin. This spectrum is subject to statistical noise because in practice one
simulates a finite number of packets. There are two sources of statistical
uncertainty in simulations like \tardis: both the number of packets in a bin,
$N$, and the set of packet luminosities $\bml = \qty{\Lum_j: j=1 \dots N}$ are
random variables.
The {total luminosity} in a bin, $\Lumtot$, is just the sum of
packet luminosities; i.e.
\begin{align}
  \lleq{gaa} \Lumtot = \sum_{j=1}^N \Lum_j \,.
\end{align}
The standard assumption is that $N$ follows a Poisson distribution with
expectation $\lambda$
\begin{align}
  \lleq{gab}
  p(N \cond \lambda) &= \frac{e^{-\lambda}\lambda^{N}}{N!},
  \quad N = 0,1,\ldots,\infty.
\end{align}
This is a good assumption if the total number of packets is large but only a
small fraction ends up in any bin. We consider all packets statistically
independent. One of the key quantities is the {single-packet luminosity
  distribution} $p(\Lum \cond \bmphi)$. The parameters $\bmphi$ govern the
distribution of luminosities and are kept general at this point. Then $\Lumtot$
follows a \emph{compound Poisson distribution} (CPD) with density
\begin{align}
  \lleq{gac}
  p(\Lumtot \cond \lambda, \bmphi) = \sum_{N=0}^\infty p(N \cond \lambda) p(\Lumtot \cond N, \bmphi),
\end{align}
where the density $p(\Lumtot \cond N, \bmphi)$ can be expressed as a
multidimensional convolution integral (see \refsec{var-lumtot}).

With the physics and all other assumptions implemented in a code like \tardis\
and the model with its input parameters $\physpar$ fixed, the outcome $\Lumtot$
always tends to the same $\Lumuniq$. It is assumed that the specific run outcome
of $\Lumtot$ depends only on the initial state (seed) of the pseudo-random
number generator. This is a necessary requirement to draw conclusions from one
simulation without having to try all other seeds. We thus assert that the
purpose of running the simulation is to learn about $\Lumuniq$ and not about
$\Lumtot$. One major difference between the two is that we can measure $\Lumtot$
directly from a given simulation whereas $\Lumuniq$ is unobservable and has to
be inferred, suggesting a Bayesian treatment. Our goal is to model what is known
about the luminosity consistently from a single simulation run even in the
extreme case with zero or one packet observed in a bin. There are several areas
in which we want to improve the state of the art. In an actual simulation, both
$\lambda$ and $\bmphi$ are not known and have to be inferred from the simulation
itself. These uncertainties should be properly taken into account. What has
previously been overlooked is that although $\Lumtot$ is an obvious estimator of
$\Lumuniq$ it is not necessarily the best. The relations are subtle, as we show
below.

We want to have sound uncertainty estimates for two reasons. First, one can stop
simulations once a desired accuracy is reached to save computing time and
electric power. Second, we do not want to discard bins with few packets as is
often done because they still contain valuable information. Third, we want to
avoid running repeated simulations just to estimate uncertainties. This is
common practice but unnecessary.

In this work we focus on radiative transfer but we stress that the same
statistical problem appears in many areas; for example in particle
physics under the name of weighted events and in Bayesian inference when
estimating a one-dimensional marginal distribution with importance sampling.

An overview of related work is presented in \refsec{rel-work}. In
\refsec{method}, we present the general results of our method. This method is
applied to a specific model for $p(\Lum \cond \bmphi)$ in \refsec{gaussian}. We
discuss the relation to previous methods and conclude the paper in
\refsec{discussion}. Derivations are presented in the Appendix \ref{sec:appendix}.

\section{Notation and related work}\label{sec:rel-work}
Suppose that we have output, or Monte Carlo data, $\data$ of one simulation
where $N$ packets with luminosities $\bml$ fall into the frequency bin under
consideration.
We define the sample mean, sample second moment, and variance as
\begin{align}
  \lleq{uaa}
  \samplel \equiv \frac{1}{N} \sum_{n=1}^N \ell_n, \quad
  \sampleltwo \equiv \frac{1}{N} \sum_{n=1}^N \ell^2_n, \quad
  \samplelvar \equiv  \sampleltwo-\samplel^2 .
\end{align}
These are estimators of the
corresponding expectation values under the single-packet luminosity distribution
given $\bmphi$. Our notation for mean and variance is
\begin{align}
  \lleq{uac}
  \mu &\equiv \expect{\Lum \cond \bmphi} \equiv \int \dd{\Lum} p(\Lum \cond \bmphi)\, \Lum,\\
  \lleq{uad}
  \sigma^2 &\equiv \variance{\Lum \cond \bmphi} \equiv \expect{\Lum^2 \cond \bmphi} - \mu^2,
\end{align}
so both $\mu$ and $\sigma$ depend on $\bmphi$.

The general approach in the literature is to identify the quantity of interest
with $\Lumtot$ and to use only the packets in one bin; i.e. information about
the luminosities from adjacent bins is ignored. A simple and fairly common
approximation \citep[e.g. ][]{2009MNRAS.398.1809K} to infer the luminosity is to
estimate a best value and errors from the Monte Carlo data as if the posterior were
\begin{align}
 \lleq{uab}
  p(\Lumtot \cond \data) = \GaussianDist\qty(\Lumtot \middle| N \samplel, N \samplel^2).
\end{align}
In words, this is a Gaussian approximation to a Poisson distribution scaled by
$\samplel$. There are several issues with this. First, there is a problem for
bins with low number of packets as it could assign nonzero probability to negative
luminosity and it breaks down entirely in the extreme case of $N=0$ packets.
Second, it ignores the uncertainty for $\bmphi$ entirely or in other words, it
supposes $p(\Lum \cond \bmphi) = p(\Lum)$ known. Third, it is symmetric in
$\Lumtot$ while the Poisson distribution is asymmetric, especially for small $N$.

The seminal papers on radiative transfer for supernovae by
\citet{1993A&A...279..447M} or photo-ionisation by \citet{2003MNRAS.340.1136E}
use the approach to run with many packets $\npack$ such that the statistical
uncertainties are negligible but do not estimate them explicitly. Specifically,
\cite{2003ASPC..288..453T} graphically compare the spectrum smoothness for
various values of $\npack$ and \cite{2004MNRAS.348.1337W} raise $\npack$ from
$10^6$ to $10^8$ packets after convergence of an iterative procedure and state
that this ``provides higher signal-to-noise''.

\cite{2015MNRAS.450..967B} compare three different techniques to follow packets
by repeating the expensive simulation 500 times on a supercomputer with
different random-number seeds and reporting the sample standard deviation on
$\Lumtot$. This approach comes at a high cost, so we seek similar uncertainty
estimates from a single run. In photon-through-dust radiative transfer,
\cite{2001ApJ...551..269G} share this objective and estimate standard errors but
their method is only applicable to their specific scenario and works reliably
only for $N \geq 10^4$ where the Poisson noise dominates over luminosity
variability. A different view on the topic is presented --- again related to
dust --- by \cite{doi:10.1111/j.1365-2966.2006.10884.x} where uncertainties are
estimated like in importance sampling. All packets can contribute in all bins
but most packets have zero contribution to a bin because they end up in another
bin. The variance estimate then underestimates the \CPD variance for small
$\npack$ and matches it only for large $\npack$.

In particle physics, the same statistical problem has been considered from a
frequentist point of view and conventionally called \emph{sum of weighted
  events}, where weight corresponds to the luminosity and an event is a packet.
\cite{BARLOW1993219} consider maximum-likelihood estimation but for weighted
events assume that $p(L\cond \bmphi)$ has negligible variance in a bin and
neglect this uncertainty when estimating asymptotic confidence intervals.
\cite{bohm_statistics_2014} explicitly consider the \CPD and suggest to
approximate it by a scaled Poisson distribution instead of a Gaussian as in
\refeq{uab}. Their goal is to estimate the variance $\variance{Q}$ for fixed
$\lambda$ whereas we consider the more common case of unknown $\lambda$.

\section{Method}\label{sec:method}

\begin{figure}
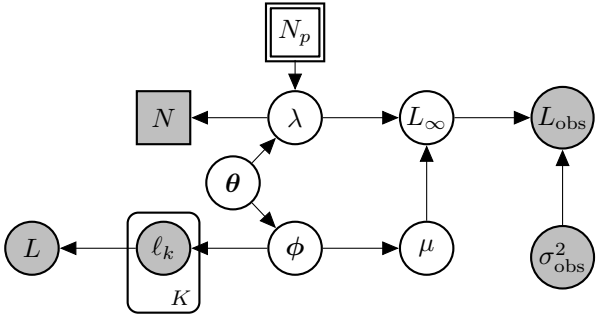

  \centering
  \tikz{ %
    \node[obsDisc] (N) {$N$} ; %
    \node[latentCont, right=of N] (lambda) {$\lambda$} ; %
    \node[latentCont, below=of lambda] (phi) {$\bmphi$};
    \node[obsCont, left=of phi] (l) {$\Lum_k$};
    \plate[] {pl} {(l)} {$K$};
    \node[obsCont, left=of l] (Q) {$\Lumtot$};
    \node[latentCont, right=of lambda] (Qinf) {$\Lumuniq$};
    \node[latentCont, right=of phi] (mu) {$\mu$};
    \edge {lambda} {N} ; %
    \edge {lambda,mu} {Qinf} ; %
    \edge {phi} {l} ; %
    \edge {phi} {mu} ; %
    \edge {l} {Q} ; %
    \node[obsCont, right=of Qinf] (Q0) {$\Lumobs$};
    \node[obsCont, below=of Q0] (varobs) {$\varobs$};
    \node[latentCont, above left=3.5mm and 3mm of phi] (theta) {$\bmtheta$};
    \edge {Qinf} {Q0};
    \edge {theta} {lambda};
    \edge {theta} {phi};
    \node[detDisc, above=4mm of lambda] (np) {$\npack$};
    \edge {np} {lambda};
    \edge {varobs} {Q0};
  }
  \caption{Representation as a Bayesian graphical model. Round/square node:
    continuous/discrete variable, plate: $K$ repetitions, white/grey background:
    (un)observed, double edge: deterministic input, single edge: random
    variable, arrow from $x \to y$: $y$ depends on $x$ probabilistically or deterministically.}
\label{fig:graphical-model}
\end{figure}

\begin{table}
  \caption{Summary of the variables used.}
  \begin{tabular}{cl}
    $\lambda, N$ & (expected) number of packets in bin; \eqref{pmb}\\
    $K \geq N$ & number of packets to infer $\bmphi$; above \eqref{vaa}\\
    $\npack$ & total number of simulated packets; \refsec{intro}\\
    $\bmtheta$ & physics parameters of interest; \eqref{saa}\\
    $\bmphi$ & fix packet luminosity distribution; \eqref{gac} and \eqref{rae}\\
    $\Lum$ & single packet luminosity;  \eqref{gaa}, \eqref{uaa}\\
    $\samplel, \mu$ & (sample) mean packet luminosity; Equations \eqref{uaa}, \eqref{uac}\\
    $\samplelvar, \sigma^2$ & (sample) variance packet luminosity; \eqref{uaa}, \eqref{uad}\\
    $\data$ & Monte Carlo data comprising $N, K, \Lum_k, \samplel, \samplelvar$\\
    $\Lumtot$ & sum of simulated packet luminosities in bin; \eqref{gaa}\\
    $\Lumuniq$ & theoretical total luminosity in bin; \eqref{vab} \\
    $\Lumobs, \varobs$ & experimental total luminosity (variance) in bin; \eqref{sab}\\
  \end{tabular}
\label{tab:notation}
\end{table}
In the existing literature, the luminosity in a bin is estimated from the Monte Carlo data $\data$ by
the sum of packet luminosities $\Lumtot$, and uncertainties are usually
quantified by (asymptotic) confidence intervals based on a normal approximation
that might be valid for large $N$. By contrast, we use $\data$ to formulate the
posterior distribution for $\Lumuniq$ directly. This allows us to easily
generalise to small $N$ and to avoid many simplifying assumptions. The more
general expressions are not much more difficult to evaluate.

To obtain accurate predictions for small $N$, it is important to use all
available information. In practice it is often the case that the luminosity
distribution depends only weakly on the frequency, if at all. Then one may
determine the luminosity distribution from packets with frequencies in and
around the bin of interest such that $K > N$ packets contribute. In the special
case where the distribution is frequency-independent one can use all packets;
i.e. $K=\npack$, in which case $K$ may be orders of magnitude larger than $N$.

We begin with basic properties of the compound Poisson distribution on
$\Lumtot$. The mean and variance are
\begin{align}
  \lleq{vaa}
  \expect{\Lumtot \cond \lambda, \bmphi} = \lambda \mu, \qquad
  \variance{\Lumtot \cond \lambda, \bmphi} = \lambda \qty(\sigma^2 + \mu^2)
\end{align}
and so depend on $\lambda$ and on the parameters $\bmphi$ that fix the
single-packet luminosity distribution $p(L \cond \bmphi)$. In a realistic
example, this distribution is not known but has to be inferred from the Monte Carlo data
$\data$.

The goal of running the simulation is to learn what the luminosity of the
physical process, $\Lumuniq$, is under given physics assumptions. If $\Lumuniq$
could be calculated analytically, then there would be no reason to run a
simulation. Hence the quantity of interest cannot be $\Lumtot$, the directly
observable outcome of the simulation. Consider increasing the total number of
packets, $\npack$, across all bins. The same total luminosity is then
distributed over more packets but the probability for a packet to end up in a
certain bin is unchanged. The Poisson parameter $\lambda$ is proportional to
$\npack$ and the mean luminosity per packet scales like $1/\npack$ such that
$\expect{\Lumtot \cond \lambda, \bmphi}$ remains unchanged. In the ideal case
$\lim_{\npack \to \infty}$, $\Lumtot$ converges to $\expect{\Lumtot \cond \lambda,
  \bmphi}$ with probability one. Our key assumption to relate $\Lumtot$ to
$\Lumuniq$ is that $\Lumuniq$ is given by the mean of $\Lumtot$; i.e.
\begin{align}
  \lleq{vab}
  \Lumuniq = \lambda \mu \,.
\end{align}
Instead of the mean, one could also take the median or the mode of the \CPD.
Asymptotically, it would make no difference; the choices correspond to different
loss functions as discussed by \cite{jaynes2003probability}. We select the mean
due to mathematical convenience. Let us suppose that in our model $\mu (\bmphi)$
is a known function and let $p(\bmphi \cond \data)$ denote the posterior for
$\bmphi$, then the sought-after posterior for $\Lumuniq$ given the Monte Carlo data is
\begin{align}
  \lleq{vac}
  p(\Lumuniq \cond\data) &= \int \dd{\bmphi}
   \GammaDist\qty(\frac{\Lumuniq}{\mu(\bmphi)} \middle| N+\half, 1) \frac{p(\bmphi \cond \data)}{\mu(\bmphi)};
\end{align}
see Appendix~\ref{sec:posterior} for definitions and the derivation. The
$\GammaDist$ distribution arises as the conjugate prior to the Poisson model and
automatically incorporates the constraint that $\Lumuniq>0$.

To highlight the connections between the variables summarised in
\reftab{notation}, \reffig{graphical-model} shows the complete model
as a Bayesian graphical model.

\begin{figure}
  \centering
  \includegraphics[width=\columnwidth]{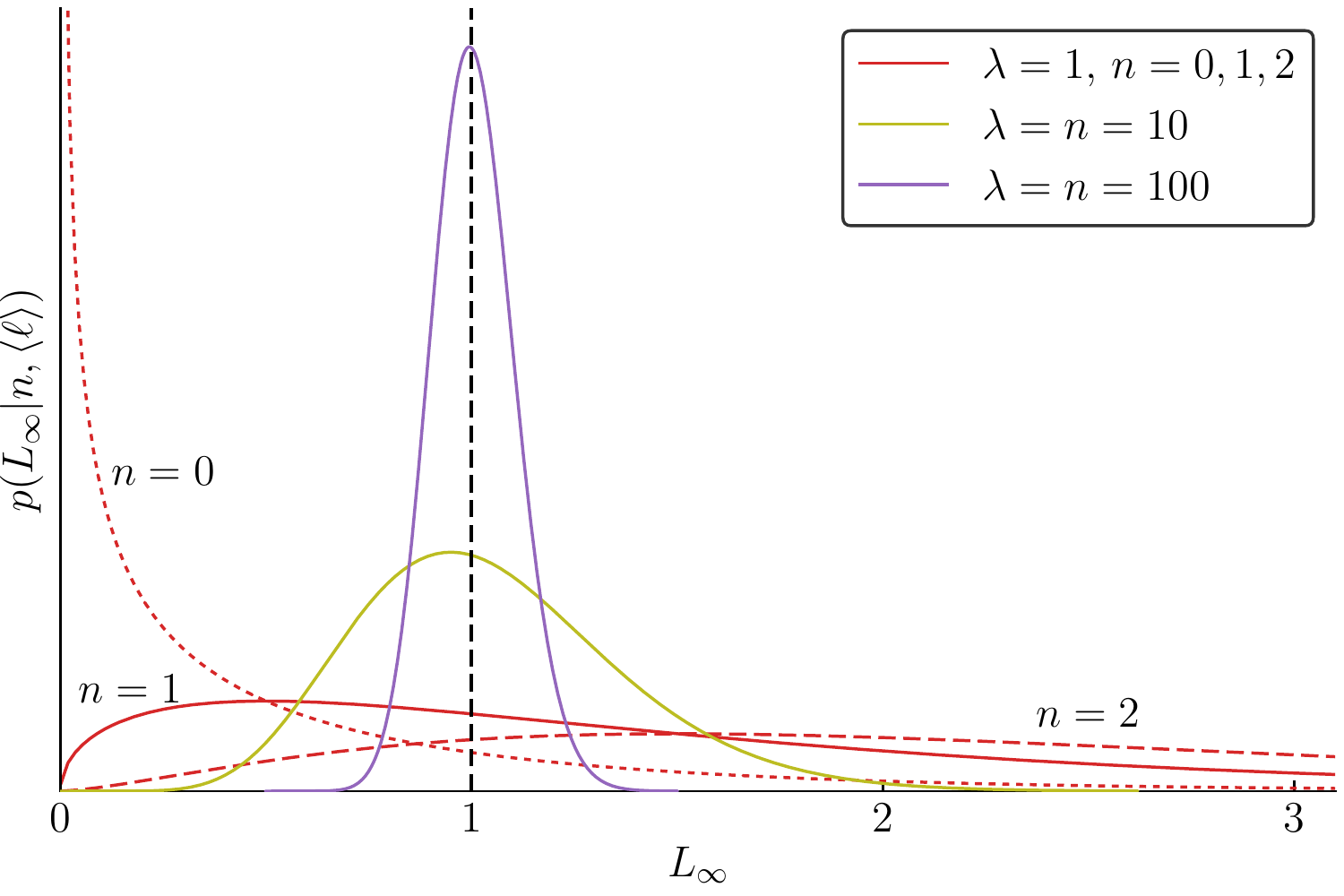}\\
  \caption{Contraction of the posterior for $\Lumuniq$ when increasing the total
    number of packets, $\npack$, in the simulation around the true value
    $\Lumuniq=1$. We set $\samplel = 1/\lambda$ in \refeq{fag} where $\lambda
    \propto \npack$.}
\label{fig:contraction}
\end{figure}

Consider the special case when the number of packets $K$ to determine $\bmphi$
is large. Then we may consider $\bmphi$ fixed, or equivalently $p(\bmphi \cond
\data)$ is a Dirac $\delta$ function, and we set $\mu$ to the sample mean from
$K$ packets, $\mu(\bmphi)=\samplel$. The Monte Carlo data are summarised by $N$ and
$\samplel$ and there is no integral to perform:
\begin{align}
  \lleq{fag}
  p(\Lumuniq \cond N, \samplel) &=                              \GammaDist\qty(\frac{\Lumuniq}{\samplel} \middle| N+\half, 1) \frac{1}{\samplel}
\end{align}
The complete solution is just a rescaled $\GammaDist$ distribution that is valid
for any $N \ge 0$, $\Lumuniq>0$, and $\samplel > 0$; see \reffig{contraction}.
For ease of comparison with the literature (\refsec{rel-work}), the posterior mean and variance  are
\begin{align}
  \lleq{vad} \expect{\Lumuniq \cond N, \samplel} &= (N+\half) \samplel,\\
  \lleq{vae} \variance{\Lumuniq \cond N, \samplel} &= (N+\half) \samplel^2.
\end{align}
as derived in \refsec{variance}. On the one hand, the variance increases with
$N$ for fixed $\samplel$. On the other hand, if $\npack$ is increased such that
$N\propto \npack$ and $\samplel \propto 1/\npack$ the variance on $\Lumuniq$
shrinks at the usual rate $1/{\npack}$ and the solution contracts around the
true value with probability one as illustrated in \reffig{contraction}.

\section{Gaussian model} \label{sec:gaussian}

For finite $K$, we have to specify the functional form $p(L \cond \bmphi)$.
Focusing on a Gaussian model that simplifies calculations helps to illuminate
the scaling behaviour. We conjecture that this model performs well in most
applications due to asymptotic normality since all we really need from $p(L
\cond \bmphi)$ to learn $\Lumuniq$ is $\mu(\bmphi)=\expect{L\cond \bmphi}$. In
contrast, modelling $\Lumtot$ is more sensitive to the higher moments of $p(L
\cond \bmphi)$.

In the Bayesian interpretation, probability is the degree of belief in the truth
of a statement conditional on all information included. In our case, the most
interesting property of $p(L \cond \bmphi)$ is its mean $\mu$. Of minor
importance is its variance $\sigma^2$ because it dictates how accurately we know
$\mu$ from the finite number of packets. The maximum-entropy distribution for
$L$ knowing the mean and variance is a Gaussian distribution, and its parameters
are just $\mu$ and $\sigma^2$ if the contribution from the $\Lum < 0$ tail of
the Gaussian is negligible, which we may safely assume if $K \gg 1$. We hence
set $\bmphi =(\mu,\sigma^2)$ and
\begin{align}
  \lleq{rae}
  p(\Lum \cond \bmphi) =
  p(\Lum \cond \mu, \sigma^2 ) &= \GaussianDist(\Lum \cond \mu, \sigma^2).
\end{align}
Using Bayes' theorem the posterior is
\begin{align}
  \lleq{ras}
  p(\mu, \sigma^2 \cond \data) = \GaussianDist(\mu\cond \samplel,\tfrac{1}{K}\sigma^2)
  \InvGammaDist\qty(\sigma^2\middle| \tfrac{K}{2},\tfrac{K}{2}\samplelvar)
\end{align}
for $\mu>0$ where we choose a conjugate Normal-$\InvGammaDist$ model with an
uninformative prior; see \refsec{app-Gaussian} for details. Now the posterior
depends on $\data$ only through $N$, $K$, and the statistics $\samplel$ and
$\samplelvar$ computed from $K$ packets. Inserting \refeq{ras} into \refeq{vac},
we have to approximate the integral using quadrature methods. We can, however,
calculate mean and variance for $\Lumuniq$ analytically:
\begin{align}
  \lleq{rat}
   \expect{\Lumuniq \cond N, K, \samplel, \samplelvar} &= (N+\half) \samplel,\\
  \lleq{rau}
  \variance{\Lumuniq \cond N, K, \samplel, \samplelvar} &= (N+\half)\qty(\frac{N+\tfrac{3}{2}}{K-2} \samplelvar+\samplel^2);
\end{align}
see \refsec{variance} for the derivation and further discussion.

\section{Discussion} \label{sec:discussion}

\subsection{Comparison to experimental data}

\begin{figure}
  \centering
  \includegraphics[width=\columnwidth]{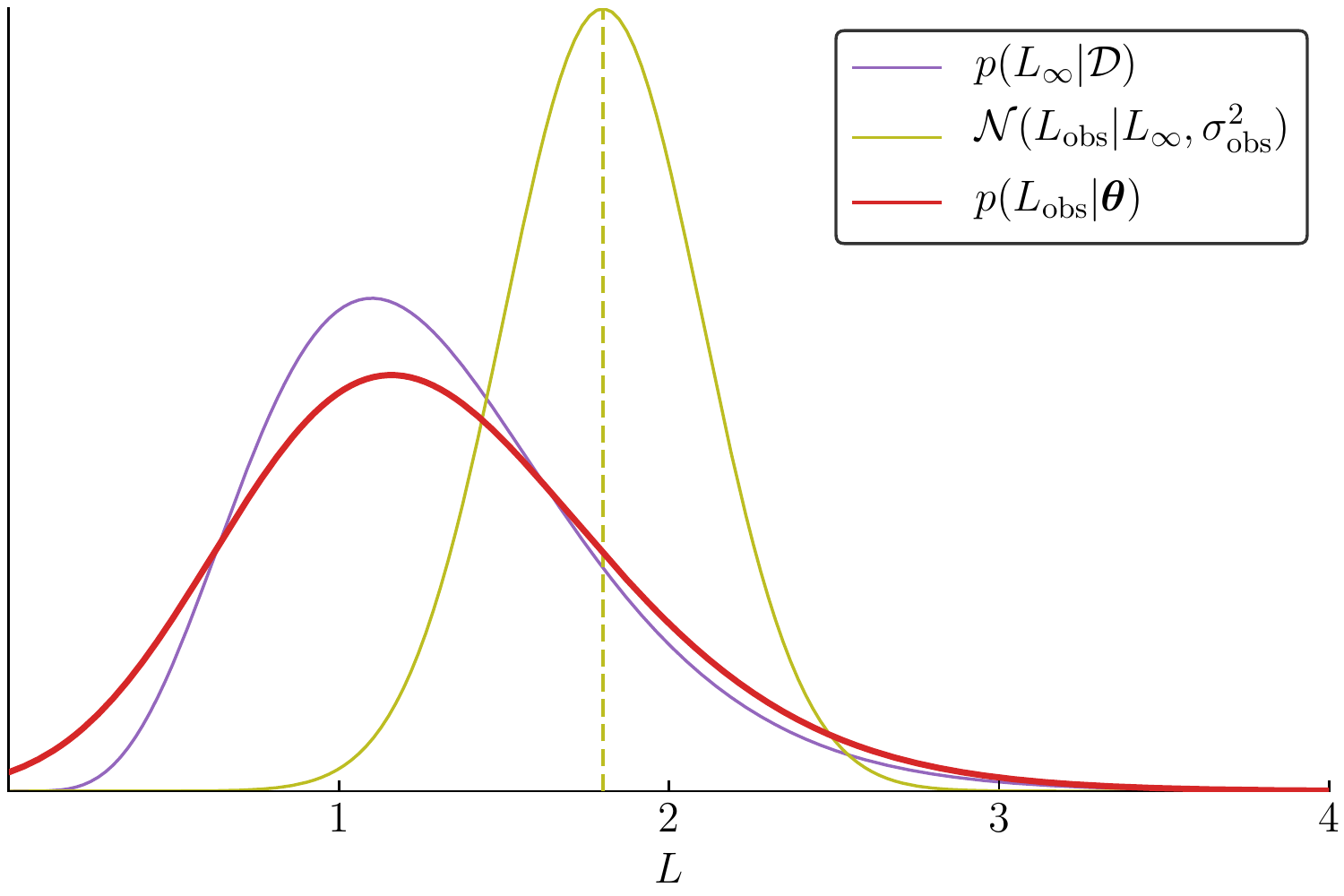}\\
  \caption{The convolution of simulation $p(\Lumuniq \cond \data)$ and
    experimental uncertainty $p(\Lumobs \cond \Lumuniq, \varobs)$ yields the
    likelihood $p(\Lumobs \cond \bmtheta)$.}
\label{fig:convolution}
\end{figure}

The ultimate goal of the presented analysis is a likelihood to compare
radiative-transfer models and experimental spectra. In a single bin, the
likelihood $\mathcal{L}(\bmtheta|\Lumobs) = p(\Lumobs|\bmtheta)$ with $\bmtheta$
denoting physics model parameters (e.g. abundances, temperatures, densities) and
$\Lumobs$ denoting the observed luminosity. This likelihood is the convolution
of the probability distribution from the radiative-transfer code (as presented
in this work) and the probability distribution for the observed luminosity (e.g.
a Gaussian distribution)
\begin{align}
  \lleq{saa}
  p(\Lumobs \cond \bmtheta) &= \int \dd{\Lumuniq} p(\Lumuniq \cond \data) p(\Lumobs \cond \Lumuniq, \bmtheta)\\
                            &= \int \dd{\Lumuniq} p(\Lumuniq \cond \data) \GaussianDist(\Lumobs \cond \Lumuniq, \varobs).
\end{align}
Note that $\bmtheta$ only implicitly affects the simulation and experimental
output in our graphical model (see \reffig{graphical-model}) and thus
does not appear explicitly in the integrand.

For simulation bins with a large number of packets (e.g. $N>25$), $p(\Lumuniq
\cond \data)$ may be accurately modelled by the Gaussian model of
\refsec{gaussian}. The $\GammaDist$ distribution in our general solution
\refeq{vac} quickly converges to a Gaussian distribution for large $N$ such that
the convolution reduces to the well known result of a Gaussian with variances
added in quadrature:
\begin{align}
 \lleq{sab}
  p(\Lumobs \cond \bmtheta) &= \GaussianDist\qty(\Lumobs \middle| (N+\half)\, \samplel, \sigma^2),
                              \intertext{where}
                              \lleq{sac}
  \sigma^2 &= \variance{\Lumuniq \middle| N, N, \samplel, \samplelvar} + \varobs,
\end{align}
where $\varobs$ is the variance of the observation model and the variance due to
simulation uncertainty is given explicitly in \refeq{rau}.

For bins with small $N$ we recommend numerical integration of \refeq{saa} using
efficient cubature techniques \citep{genz1980remarks,cubature}.
\reffig{convolution} illustrates the effect of the convolution: the resulting
distribution $p(\Lumobs \cond \bmtheta)$ is smeared out taking into account both
simulation and experimental uncertainty. In \reffig{convolution}, we choose the
$\GammaDist$ distribution of \refeq{fag}, which requires $K \gg 1$, for $N=6$
and $\samplel=0.2$. In this case, the variance due to simulation from
\refeq{vae} is 0.26 and dominates the experimental uncertainty
$\varobs=0.3^2=0.09$.

In a Bayesian analysis, one would evaluate the presented likelihood for each
frequency bin in every iteration of, for example, a Markov chain over $\bmtheta$
given the observed value $\Lumobs$ as indicated by the dashed vertical line in
\reffig{convolution}. For $\Lumobs$ or $\Lumuniq$ close to zero, it is necessary
to manually normalise the outcome to ensure that $\int_{\Lumobs>0} \dd{\Lumobs}
p(\Lumobs \cond \bmtheta) =1$ because the convolution with the Gaussian
$\GaussianDist(\Lumobs \cond \Lumuniq, \varobs)$ may shift some probability mass
into the unphysical region $\Lumobs < 0$. This effect is relevant even for
parameter inference because the normalisation depends on $\bmtheta$. For the
analytic result \refeq{sab}, the correction can be done with the cumulative
of the Gaussian distribution. For the numerical integration, we evaluate
$p(\Lumobs \cond \bmtheta)$ on a grid of $\Lumobs$ values instead of just the
observed value, and then use, for example, Simpson's quadrature rule to
normalise. In the example of \reffig{convolution}, the normalisation was off by
just 0.5 \% function but upon increasing $\varobs$ from $0.3^2 \to 1$, the
normalisation is off by 12 \%.

A simple example illustrates the potential computational savings of
our method. Let us assume a reasonable relative experimental uncertainty
$\sigobs/\Lumobs = 5 \%$. To achieve the same level of precision on $\Lumuniq$
in a typical \tardis{} run with $\samplel=1.41, \samplelvar=0.001$, we solve
\begin{align}
  \lleq{sad}
  \frac{\sqrt{\variance{\Lumuniq \middle| N, N, \samplel, \samplelvar}}}{\expect{\Lumuniq \cond N, N, \samplel, \samplelvar}} = 0.05
\end{align}
to find about $N=400$ samples in a bin are needed. Let us consider the
simulation uncertainty to be irrelevant at the 1~\% level. Then we would need
$N=10000$, or 25 times more packets in the simulation.

\subsection{Aspects of the derivation}

The posterior means in \refeq{vad} and \refeq{rat} agree exactly and the
variance \refeq{vae} is obtained from \refeq{rau} in the limit $N \ll K$. This
is a good consistency check because the derivation of \refeq{vae} assumed $K \to
\infty$ from the start. In the case $1 \ll N \ll K$, mean and variance agree
with the normal approximation in \refeq{uab} used previously when $N+\half
\approx N$. The extra $+\half$ is due to our usage of the Jeffreys prior for
$\lambda$; it leads to a reduced bias and reduced variance compared to a uniform
prior that would contribute an extra $+1$, see \refsec{variance} for further
details. In contrast to the normal approximation, our posterior $p(\Lumuniq
\cond \data)$ is applicable for any $N$, even $N=0$. In that case, $K>2$ is
required for finite variance and the larger $K$ is, the better. It is evident
from the extra term $\propto \samplelvar$ in \refeq{rau} that including the
sampling uncertainty of $\lum$ increases the variance on $\Lumuniq$, so the
asymptotic approximation \refeq{vae} and the very similar literature result
\ref{uab} may severely underestimate the total uncertainty on $\Lumuniq$ if
$\sqrt{\samplelvar}/\samplel$ is of $\order{1}$, which occurs, for example, if
$p(\lum \cond \bmphi)$ peaks near $\lum = 0$.

\subsection{Comparison to the compound Poisson}

We stress again that we formulate a posterior for $\Lumuniq$ and we do not model
the \CPD for $\Lumtot$ as for example \cite{bohm_statistics_2014} do. We
could also model the posterior predictive distribution for $\Lumtot$ given the
simulation output, $p(\Lumtot \cond \data)$. Although there is no fundamental
problem, there are many more integrals to solve --- in particular for small $N$
--- instead of the single one in \refeq{vac}, and the choice of $p(L \cond
\bmphi)$ is crucial to getting a tractable result at all. We present the
derivation in \refsec{var-lumtot} that shows the same problems also hold true in
a frequentist analysis. The conclusion from the posterior predictive is that the
dominant contribution to the variance in the Bayesian case is $\propto 2n$, or
just twice as large as in our central results of Equations \ref{vae} and
\ref{rau} because there is Poisson uncertainty from the one observation of $N$
and the same uncertainty again in the prediction of a future output of the
compound Poisson process. Another difference is that $p(\Lumuniq \cond \data)$
is unimodal but $p(\Lumtot \cond \data)$ is multimodal due to superposition as
in \refeq{gac}, although the impact can be made small by suitable data
preprocessing.

Coincidentally, someone who mistakes $\Lumtot$ for $\Lumuniq$ and sets
$\expect{\Lum \cond \bmphi} = \samplel$ ignoring all uncertainties except due to
the Poisson process asymptotically gets very similar mean and variance as with
our more general method; compare \refeq{uab} to Equations \ref{vad} and
\ref{vae}. But our method also applies for small $N$ and includes all
uncertainties. In special cases, all integrals can even be done analytically.

\section{Conclusion}

In conclusion, our new method to treat statistical uncertainties in the
estimation of the luminosity in a frequency bin based on packets from a Monte
Carlo radiative-transfer simulation is superior to previous approaches in
several ways. First, we present a proper posterior distribution that is easy to
evaluate for any value of the luminosity, not just an error estimate. Second,
our posterior is valid for large and small $N$, even for $N=0$ packets. Thus no
bin should ever be discarded as each contains valuable information. Knowledge of
the uncertainties allows to stop the simulation once a certain threshold is
reached.

Future directions of research include a fundamental generalisation from
independent bins and the Poisson model to bins coupled in a multinomial model.
One would then predict the luminosities in all bins simultaneously. In a single
bin, the Poisson rate $\lambda$ would correspond to $ c\, \npack$, where $c$ is the probability for
a packet to end up in this bin. The sum of those probabilities is one, a
constraint that we do not use in our current formulation. Additional useful
constraints include modelling $\mu$ as frequency-dependent, for example in a
smooth fashion with a Gaussian process as in \cite{rasmussen2006cki} that, once
trained, yields a Gaussian distribution for $\mu$ at every bin centre. Then the
multi-dimensional integral over $\bmphi$ in \refeq{vac} reduces to the simpler
one-dimensional integral \refeq{fah} and the mean would be known very precisely
in regions with many packets. The disadvantages include the extra computing effort
to train the Gaussian process in every iteration of a fit, sensitivity to the
hyperparameters of the Gaussian process, and an extra level of complexity where
user attention is needed.
One reason why $\mu$ could vary with the frequency is that the probability of
scattering say a single photon on an iron atom depends on the frequency of the
photon so the change of luminosity of a photon packets also depends on the
frequency. Including extra constraints should lead to a tighter posterior on
$\mu$ but it is not clear if tractable results can be obtained, and presumably
they are not as easy to compute as in our current approach.

\section*{Acknowledgements}

HCE gratefully acknowledges support by the Alexander von Humboldt Foundation,
the Excellence Cluster Universe, and the South African National Research
Foundation. FB thanks the National Institute of Theoretical Physics and
Stellenbosch University for hospitality during preparation of this manuscript.
WEK acknowledges the support through an ESO Fellowship. We would also like to
thank Allen Caldwell for thoughtful discussion and suggested edits of the
manuscript. The anonymous reviewer helped greatly in clarifying our exposition.

\bibliographystyle{mnras}
\bibliography{references}

\section{Appendix} \label{sec:appendix}

\subsection{Derivation of posterior for $\Lumuniq$} \label{sec:posterior}
The first step is the posterior for $\lambda$ given $N$, the number of packets
with a frequency in the bin under consideration.  Bayes' Theorem yields
\begin{align}
  \lleq{pmc}
  p(\lambda\cond N) %
  &= \frac{p(N\cond\lambda)\,p(\lambda)} {p(N)}
  \ =\ \frac{p(N\cond\lambda)\,p(\lambda)} %
  {\int \rmdx{\lambda} p(N\cond\lambda)\,p(\lambda)}
\end{align}
with the Poisson distribution
\begin{align}
  \lleq{pmb}
  p(N\cond\lambda) &= \frac{e^{-\lambda}\lambda^{N}}{N!}
  \quad N = 0,1,\ldots,\infty.
\end{align}
The conjugate prior for $\lambda$ is the $\GammaDist$ distribution with shape
parameter $\alpha > 0$ and rate parameter $\beta > 0$ defined as
\begin{align}
  \lleq{gma}
  \GammaDist(x\cond \alpha, \beta)
  &= \frac{\beta^{\alpha}}{\Gamma(\alpha)} x^{\alpha-1} e^{-\beta x}\,,
    \quad 0 \leq x < \infty.
\end{align}
With initial hyperparameters $\alpha_0$ and $\beta_0$, the posterior is
\begin{align}
  \lleq{pms}
  p(\lambda\cond N)
  &= \GammaDist(\lambda \cond \alpha = N+\alpha_0,\, \beta = 1+\beta_0).
\end{align}
We set $\beta_0 = 0$, which implies effectively no observations in the prior.
For $\alpha_0$, we consider three options: $\alpha_0=1$ is a uniform prior,
$\alpha_0=\half$ is the prior advocated by \cite{Jeffreys453}, which coincides
with the reference prior by \cite{bernardo_reference_1979}, and $\alpha_0=0$ is
the transformation-group prior described by \cite{jaynes2003probability}. The
Jaynes prior is derived assuming $\lambda \propto$ measurement time and that any
rescaling of time cannot change the state of knowledge. This prior is degenerate
for an observation $N=0$ and thus not useful for us because it is common to have
empty bins. We prefer the Jeffreys prior over the uniform prior because the
former corresponds to a state of knowledge with minimum expected impact on the
posterior and carries the same information in any one-to-one transformation of
parameters, thus we set $\alpha_0=\half$.

The second step is to choose a specific distribution for the luminosity of a
single packet that comes out of the simulation, $p(L \cond \bmphi)$. This is
problem dependent and we leave this decision unspecified. In any case, the
posterior for $\bmphi$ has the generic structure
\begin{align}
  \lleq{pmt}
  p(\bmphi \cond \data) \propto  p(\bmphi) \prod_{n=1}^N p(\Lum_n \cond \bmphi).
\end{align}
We assume that the luminosity distribution permits to calculate the mean
luminosity $\mu(\bmphi)$ analytically.
Using the basic laws of probability, we can propagate the uncertainty on
$\lambda$ and $\bmphi$ to arrive at the desired posterior for $\Lumuniq =
\lambda \mu(\bmphi)$ using the Dirac $\delta$ function $p(\Lumuniq \cond
\lambda, \bmphi) = \delta(\Lumuniq - \lambda \mu(\bmphi))$:
\begin{align}
  \lleq{fae}
  p(\Lumuniq \cond \data)
  &= \int \rmdx{\lambda} \rmdx{\bmphi} p(\Lumuniq \cond \lambda, \bmphi, \data) p(\lambda, \bmphi \cond \data)\\
  &= \int \rmdx{\lambda} \rmdx{\bmphi} p(\Lumuniq \cond \lambda, \bmphi) p(\lambda \cond N) p(\bmphi \cond \data)\\
  &= \int \rmdx{\lambda} \rmdx{\bmphi} \delta(\Lumuniq - \lambda \mu(\bmphi)) \nonumber\\
  &  \quad \times \GammaDist(\lambda \cond N+\half, 1) p(\bmphi \cond \data).
\end{align}
Performing the integral over $\lambda$ with the help of the $\delta$ function,
we arrive the general result
\begin{align}
  \lleq{faf}
  p(\Lumuniq \cond \data) &= \int \rmdx{\bmphi}
   \GammaDist\qty(\frac{\Lumuniq}{\mu(\bmphi)} \middle| N+\half, 1) \frac{p(\bmphi \cond \data)}{\mu(\bmphi)}.
\end{align}
Formally we can rewrite the result as a one-dimensional integral because $\bmphi$
appears only through $\mu$
\begin{align}
  \lleq{fah}
  p(\Lumuniq \cond \data)
  &= \int \rmdx{\mu} \GammaDist\qty(\frac{\Lumuniq}{\mu} \middle| N+\half, 1) \frac{p(\mu \cond \data)}{\mu}
\end{align}
but $p(\mu \cond \data)$ requires a usually intractable integral over $\bmphi$.

\subsection{Details on the Gaussian model} \label{sec:app-Gaussian}

The luminosity distribution $p(L \cond \bmphi)$ (see \refeq{rae}) can be
inferred from Monte Carlo data $\data=(K, \bml)$ with $K$ samples, where $K = N$ means
packets from just one bin are used. For weak dependence on the frequency, one
may use several adjacent bins such that $K>N$. Under the assumption that $p(L
\cond \bmphi)$ is independent of the frequency, one can use all packets; i.e.
$K = \npack$. Assuming independent packets, the posterior is
\begin{align}
  \lleq{rac}
  p(\bmphi \cond \data) &=
  p(\mu,\sigma^2\cond K,\bml)\\
  &= \frac{
    \prod_{k=1}^K \GaussianDist(\ell_k \cond\mu,\sigma^2)\;
    p(\mu)\;
    p(\sigma^2)}
  {p(\data)} \, .
\end{align}
In keeping with prior information that luminosities are always positive, we
assign a uniform prior $p(\mu\cond\mu_{\rm max}) = 1/\mu_{\rm max}$ independent
of $N$; as before, the posterior is independent of $\mu_{\rm max}$ if chosen
large enough. For $\sigma^2$, we use an Inverse Gamma prior with hyperparameters
$(a_0,b_0)$,
\begin{align}
  \lleq{rad}
  p(\sigma^2\cond a_0,b_0)
  &= \InvGammaDist(\sigma^2\cond a_0,b_0)\\
  &= \frac{\qty(b_0)^{a_0}}{\Gamma(a_0)} (\sigma^2)^{-a_0-1} \exp(-b_0/\sigma^2)
\end{align}
which is ``conjugate'' to the normal distribution likelihood of
\refeq{rae} in that the posterior for $(\mu,\sigma^2)$ is also a
Gaussian in $\mu$ and an Inverse Gamma in $\sigma^2$,
\begin{align}
  \lleq{rbd}
  p(\mu,\sigma^2\cond K,\bml,a_0,b_0)
  &= \GaussianDist(\mu\cond \mu_K,\sigma_K^2)\nonumber\\
  &\quad \times \InvGammaDist(\sigma^2\cond a_K,b_K)
\end{align}
if the contribution from $\mu<0$ can be neglected. The hyperparameters are
updated by the Monte Carlo data,
\begin{align}
  \mu_K &= \samplel, \lleq{rbe}\\
  \sigma_K^2 &= \frac{\sigma^2}{K},  \lleq{rbf}\\
  a_K &= a_0 + \frac{K}{2},  \lleq{rbg}\\
  b_K &= b_0 + \frac{K}{2}\samplelvar.   \lleq{rbh}
\end{align}
Results for small $K$ are inaccurate so one would typically pool more packets.
In that case $K \gg 1$, the prior is less important and we may simply choose
$a_0 = b_0 = 0$ for a noninformative prior on $\sigma^2$.

\subsection{Variance for $\Lumuniq$} \label{sec:variance}

Retaining the hyperparameter $\alpha_0$ of the prior $p(\lambda \cond \alpha_0)$ (see
\refeq{pms}), the moments of the posterior $p(\Lumuniq \cond \data)$ of
\refeq{faf} for the asymptotic normal-inverse-gamma model where $ \bmphi = \mu,
\sigma^2$ (see \refeq{rbd}) are
\begin{align}
  \lleq{taa}
  \expect{\Lumuniq^r \cond \data} =
  & \int \dd{\Lumuniq} \dd{\mu} \dd{\sigma^2} \frac{\Lumuniq^r}{\mu}\\
  & \times \GammaDist\qty(\frac{\Lumuniq}{\mu} \cond N+\alpha_0,1) \nonumber\\
  & \times \mathcal{N}\qty(\mu \cond \mu_K, \sigma^2_K) \InvGammaDist(\sigma^2\cond a_K,b_K) \,. \nonumber
\end{align}
Using the definitions of $\mu_K, a_K, b_K$ in Equations \ref{rbe}--\ref{rbh} and
the known first and second moments of the normal and inverse-gamma distributions, we
find after some algebra
\begin{align}
  \lleq{tab}  \expect{\Lumuniq \cond \data} &= (N+\alpha_0) \samplel,\\
  \lleq{tac} \expect{\Lumuniq^2 \cond \data} &= (N+\alpha_0)(N+\alpha_0+1)\qty(\samplel^2 + \frac{\samplelvar}{K-2}),
\end{align}
where $\samplel$ and $\samplelvar$ are based on $K$ packets. From this the
variance follows as
\begin{align}
  \lleq{tai}
  \variance{\Lumuniq \cond \data} &= (N+\alpha_0)\qty(\frac{N+\alpha_0+1}{K-2} \samplelvar+\samplel^2).
\end{align}
Now let us focus on the special case with many packets in a bin; i.e. $N=K$ and $N \gg 1$:
\begin{align}
  \lleq{tad}
  \variance{\Lumuniq \cond \data} = N \sampleltwo + \qty[(2\alpha_0+1) \sampleltwo - (\alpha_0+1) \samplel^2] + \order{\tfrac{1}{N}}.
\end{align}
where we set $N/(N-2) \to 1$. The three groups of terms allow an intuitive
explanation. The dominant term $N\sampleltwo$ captures the Poisson uncertainty
inherent in the \CPD. Ignoring the other terms, we see that if we average over
the data distribution for fixed $\lambda$ and $\bmphi$, we obtain the
variance for the \CPD given in \refeq{vaa}:
\begin{align}
  \lleq{tah}
  &\expect{\variance{\Lumuniq \cond \data} \middle| \lambda, \bmphi} \nonumber\\
  &= \sum_N \int \dd{\bml} p(N \cond \lambda) p\qty(\bml \cond \bmphi) N \frac{1}{N} \sum_{n=1}^N \ell_n^2\\
  &= \sum_N p(N \cond \lambda) N \expect{\Lum^2 \cond \bmphi}\\
  &= \lambda \expect{\Lum^2\cond \bmphi} = \lambda (\mu^2 + \sigma^2) = \variance{\Lumtot \cond \lambda, \bmphi}.
\end{align}
In words, in the asymptotic regime our method that assumes $\lambda$ and
$\bmphi$ unknown on average yields a variance on the posterior for $\Lumuniq$
that is equal to the variance of the \CPD for $\Lumtot$ where $\lambda$ and
$\bmphi$ are assumed known! So with enough samples, the variance of the
luminosities is irrelevant --- there are $N$ samples to determine the moments
--- and only the Poisson uncertainty matters because it is inferred from a
single observation.

The cross-terms in \refeq{tad} that are independent of $N$ are due to the
simultaneous uncertainty of $\lambda$ and $\mu, \sigma^2$. Finally, the terms of
$\order{1/N}$ follow the usual law for inferring mean and variance from $N$
identically distributed samples.

The cross-terms show the quantitative effect of the hyperparameter $\alpha_0$ in
the prior for $\lambda, p(\lambda \cond \alpha_0)$. The flat prior
($\alpha_0=1$) yields the largest variance, the Jaynes prior ($\alpha_0=0$)
yields the smallest, and Jeffreys prior ($\alpha_0=1/2$) leads to something in
between:
\begin{align}
  \lleq{tar}
  (2\alpha_0+1) \sampleltwo - (\alpha_0+1) \samplel^2=
  \begin{cases}
    3 \samplelvar + \samplel^2, & \text{flat}\\
    2 \samplelvar + 1/2 \samplel^2, & \text{Jeffreys} \,.\\
    \samplelvar, & \text{Jaynes}\\
  \end{cases}
\end{align}
Repeating the calculation for fixed $\bmphi$, or $K \to \infty$, as in
\refeq{fag}, the $r$-th moment is
\begin{align}
  \lleq{tat}
  \expect{\Lumuniq^r \cond \data} =
                    \int \dd{\Lumuniq}  \frac{\Lumuniq^r}{\mu} \GammaDist\qty(\frac{\Lumuniq}{\mu} \cond N+\alpha_0,1) \nonumber\\
\end{align}
and we obtain the same mean as in \refeq{tab} and the variance
\begin{align}
  \lleq{taq}
  \variance{\Lumuniq \cond \data} = (N+\alpha_0) \samplel^2,
\end{align}
which is the limit of \refeq{tai} as $K \to \infty$. Interestingly, now the
variance only depends on the first moment whereas in \refeq{tad} it is
$\propto \sampleltwo$ for large $N$. What seems like a paradox
%
%
is a direct consequence of probability theory as an extension of logic. If we
assume from the start that $\expect{L}$ is known, then all other properties of
$p(L \cond \bmphi)$ are irrelevant with regard to inferring $\Lumuniq$. If on
the contrary, we have to infer the mean from the Monte Carlo data, then it matters how large
the variance of $p(L \cond \bmphi)$ is because the smaller the variance, the
better the mean can be inferred from a fixed number of samples. Since in
practice we do not know $\bmphi$, we recommend \refeq{tai} to estimate
$\variance{\Lumuniq \cond \data}$. This follows the recommendation by
\cite{jaynes2003probability} to perform calculations for finite $N$ and to take
the limit $N \to \infty$ only in the very end to avoid a paradox.

\subsection{Variance for $\Lumtot$} \label{sec:var-lumtot}

Repeating \refeq{gac}, the basic definition of the \CPD is
\begin{align}
  \lleq{taj}
  p(\Lumtot \cond \lambda, \bmphi) = \sum_{N=0}^\infty p(N \cond \lambda)\; p(\Lumtot \cond N, \bmphi).
\end{align}
To obtain the posterior predictive given $\data$, we have to average over the
posterior for $\lambda$ and $\bmphi$ given in Equations \ref{pms} and \ref{rac}:
\begin{align}
  \lleq{tak}
  p(\Lumtot \cond \data) = \int \dd{\lambda} \dd{\bmphi}\;  p(\Lumtot \cond \lambda, \bmphi)\;  p(\lambda \cond \data)\;  p(\bmphi \cond \data).
\end{align}
The second term on the right in \refeq{taj} can be expressed as a convolution
upon expanding with the $N$ packet luminosities $\bml$
\begin{align}
  \lleq{tal}
  p(\Lumtot \cond N, \bmphi) &=  \int \rmdx{\bml}\;
    p(\Lumtot\cond N,\bml)\; p(\bml \cond N, \bmphi)\\
  &= \int \rmdx{\bml}\; \delta(\Lumtot - \textstyle\sum_{n=1}^N \Lum_n)\; \prod_{n=1}^N p(\Lum_n\cond \bmphi).
\end{align}
To get a tractable expression, it is convenient to choose $p(\Lum \cond \bmphi)$
such that the integral in \refeq{tal} can be solved analytically. This occurs
for stable distributions for which the sum of luminosities follows the same
distribution as each individual luminosity, although with different $\bmphi$.
Examples include the Gaussian model of \refsec{app-Gaussian} where $\Lumtot \sim
\GaussianDist\qty(N \mu, N \sigma^2)$ or a $\GammaDist$ model where $p(\Lum
\cond \bmphi) = \GammaDist(\Lum \cond \alpha, \beta)$ and $\Lumtot \sim
\GammaDist(N \alpha, \beta)$. The latter has the advantage that it automatically
excludes $\Lum < 0$. In either case, for $\lambda$ large, the central limit
theorem for the \CPD holds, which eliminates both $\sum_N$ and $\int \dd{\bml}$; see \citet[Theorem 4.3.1]{bening2002generalized}:
\begin{align}
  \lleq{tam}
  p(\Lumtot \cond \lambda,\bmphi)
  &= \GaussianDist \qty( \Lumtot \cond \lambda \mu, \lambda (\mu^2 {+} \sigma^2) ).
\end{align}
Repeating the calculation for the moments with this Gaussian approximation for
the posterior predictive for $\Lumtot$,
\begin{align}
  \lleq{tan}
  \expect{\Lumtot^r \cond \data} =
  &\int \dd{\Lumtot} \dd{\lambda}\dd{\mu} \dd{\sigma^2} \Lumtot^r \GaussianDist \qty( \Lumtot \cond \lambda \mu, \lambda (\mu^2 {+} \sigma^2) ) \nonumber \\
  & \times \GammaDist\qty(\lambda \cond N+\alpha_0,1) \\
  & \times \mathcal{N}\qty(\mu \cond \mu_K, \sigma^2_K) \InvGammaDist(\sigma^2\cond a_K,b_K) \,. \nonumber
\end{align}
we find
\begin{align}
  \lleq{tae} \expect{\Lumtot \cond \data} &= (N+\alpha_0) \samplel,\\
  \lleq{tas} \expect{\Lumtot^2 \cond \data} &= (N+\alpha_0) \\
  & \quad \times \bigg\{ (N+\alpha_0+2) \qty(\samplel^2 + \frac{\samplelvar}{K-2}) \nonumber
                               + \frac{\samplelvar}{1-2/K} \bigg\}.
\end{align}
Again setting $N=K$ and $N/(N-2) \to 1$ for large $N$ yields 
\begin{align}
 \lleq{taf}
  \variance{\Lumtot \cond \data} = 2n \sampleltwo + \qty[(3\alpha_0-2) \sampleltwo - (8-\alpha_0) \samplel^2] + \order{\tfrac{1}{N}}.
\end{align}

\end{document}

%% file: symbols.tex
\newcommand{\bml}{{\bm{\ell}}}

\newcommand{\bmphi}{{\bm{\phi}}}

\newcommand{\bmtheta}{{\bm{\theta}}}